\documentclass[aps,prc,twocolumn,superscriptaddress,floatfix]{revtex4-1}
\usepackage{textcomp}
\usepackage{nicefrac}
\usepackage{longtable}
\usepackage{graphicx}
\usepackage{multirow}
\usepackage{amsmath}
\usepackage{url}

\begin{document}
 
 \title{Re-examining the $^{26}$Mg($\alpha,\alpha^\prime$)$^{26}$Mg reaction - probing astrophysically important states in $^{26}$Mg.}
 
 \author{P. Adsley}
 \email{padsley@gmail.com}
 \affiliation{Department of Physics,  Stellenbosch University, Private Bag X1, 7602 Matieland, Stellenbosch, South Africa}
 \affiliation{iThemba Laboratory for Accelerator Based Sciences, Somerset West 7129, South Africa}
 \affiliation{Institut de Physique Nucl\'{e}aire d'Orsay, UMR8608, CNRS-IN2P3, Universit\'{e} Paris Sud 11, 91406 Orsay, France}
  
 \author{J.W. Br\"{u}mmer}
 \author{K.C.W. Li}
  \affiliation{Department of Physics,  Stellenbosch University, Private Bag X1, 7602 Matieland, Stellenbosch, South Africa}
 
 \author{D.J. Mar\'{i}n-L\'{a}mbarri}
 \affiliation{iThemba Laboratory for Accelerator Based Sciences, Somerset West 7129, South Africa}
 \affiliation{Department of Physics, University of the Western Cape, P/B X17, Bellville 7535, South Africa} 
 \affiliation{Instituto de F\'{\i}sica, Universidad Nacional Aut\'{o}noma de M\'{e}xico, Apdo. Postal 20-364, 01000 Cd. M\'{e}xico, Mexico}
 \author{N.Y. Kheswa}
 \affiliation{iThemba Laboratory for Accelerator Based Sciences, Somerset West 7129, South Africa}
 \author{L.M. Donaldson}
  \affiliation{iThemba Laboratory for Accelerator Based Sciences, Somerset West 7129, South Africa}
 \affiliation{School of Physics, University of the Witwatersrand, Johannesburg 2050, South Africa}
   \author{R. Neveling}
 \affiliation{iThemba Laboratory for Accelerator Based Sciences, Somerset West 7129, South Africa}
 \author{P. Papka}
  \affiliation{Department of Physics,  Stellenbosch University, Private Bag X1, 7602 Matieland, Stellenbosch, South Africa}
 \affiliation{iThemba Laboratory for Accelerator Based Sciences, Somerset West 7129, South Africa}
 \author{L. Pellegri}
 \affiliation{iThemba Laboratory for Accelerator Based Sciences, Somerset West 7129, South Africa}
 \affiliation{School of Physics, University of the Witwatersrand, Johannesburg 2050, South Africa}
 \author{V. Pesudo}
 \affiliation{iThemba Laboratory for Accelerator Based Sciences, Somerset West 7129, South Africa}
 \affiliation{Department of Physics, University of the Western Cape, P/B X17, Bellville 7535, South Africa}
 \author{L.C. Pool}
 \affiliation{iThemba Laboratory for Accelerator Based Sciences, Somerset West 7129, South Africa}
 \author{F.D. Smit}
 \affiliation{iThemba Laboratory for Accelerator Based Sciences, Somerset West 7129, South Africa}
 \author{J.J. van Zyl}
  \affiliation{Department of Physics,  Stellenbosch University, Private Bag X1, 7602 Matieland, Stellenbosch, South Africa}

 \date{\today}
 
\begin{abstract}

\begin{description}
\item[Background] The $^{22}$Ne($\alpha,n$)$^{25}$Mg reaction is one of the neutron sources for the $s$-process in massive stars. The properties of levels in $^{26}$Mg above the $\alpha$-particle threshold control the strengths of the $^{22}$Ne($\alpha,n$)$^{25}$Mg and $^{22}$Ne($\alpha,\gamma$)$^{26}$Mg reactions. The strengths of these reactions as functions of temperature are one of the major uncertainties in the $s$-process.
\item[Purpose] Information on the existence, spin and parity of levels in $^{26}$Mg can assist in constraining the strengths of the $^{22}$Ne($\alpha,\gamma$)$^{26}$Mg and $^{22}$Ne($\alpha,n$)$^{25}$Mg reactions, and therefore in constraining $s$-process abundances.
\item[Methods] Inelastically scattered $\alpha$ particles from a $^{26}$Mg target were momentum-analysed in the K600 magnetic spectrometer at iThemba LABS, South Africa. The differential cross sections of states were deduced from the focal-plane trajectory of the scattered $\alpha$ particles. Based on the differential cross sections, spin and parity assignments to states are made.
\item[Results] A newly assigned $0^+$ state was observed in addition to a number of other states, some of which can be associated with states observed in other experiments. Some of the deduced $J^\pi$ values of the states observed in the present study show discrepancies with those assigned in a similar experiment performed at RCNP Osaka. The reassignments and additions of the various states can strongly affect the reaction rate at low temperatures.
\item[Conclusion] The number, location and assignment of levels in $^{26}$Mg which may contribute to the $^{22}$Ne$+\alpha$ reactions are not clear. Future experimental investigations of $^{26}$Mg must have an extremely good energy resolution in order to separate the contributions from different levels. Coincidence investigations of $^{26}$Mg provide a possible route for future investigations.
\end{description}

\end{abstract}

\maketitle

\section{Astrophysical Background}

The slow neutron-capture process ($s$-process) is responsible for the synthesis of about half of the overall inventory of elements heavier than iron \cite{RevModPhys.83.157}. Two nuclear reactions contribute most of the neutrons to the $s$-process: $^{13}$C($\alpha,n$)$^{16}$O and $^{22}$Ne($\alpha,n$)$^{25}$Mg. The $^{22}$Ne($\alpha,n$)$^{25}$Mg reaction contributes to the main component of the $s$-process during thermal pulses in low- and intermediate-mass asymptotic giant branch (AGB) stars \cite{0004-637X-497-1-388}, and contributes to the weak branch of the $s$-process in massive stars during helium burning \cite{1991ApJ...367..228R} and carbon-shell burning \cite{1991ApJ...371..665R}.
The efficacy of the $^{22}$Ne($\alpha,n$)$^{25}$Mg reaction as a neutron source depends on the strengths of the $^{22}$Ne($\alpha,n$)$^{25}$Mg and $^{22}$Ne($\alpha,\gamma$)$^{26}$Mg reactions. The $^{22}$Ne($\alpha,n$)$^{25}$Mg reaction is slightly endothermic ($Q = -478.29$ keV) and thus does not operate until higher temperatures (approximately 0.3 GK) are reached. Meanwhile, the $^{22}$Ne($\alpha,\gamma$)$^{26}$Mg reaction ($Q = 10.615$ MeV) can continually operate, depleting the available inventory of $^{22}$Ne and thereby reducing the total neutron exposure. In order to constrain the production of $s$-process nuclides, it is important to know the $^{22}$Ne($\alpha,n$)$^{25}$Mg and $^{22}$Ne($\alpha,\gamma$)$^{26}$Mg reaction rates over a range of temperatures.

Owing to the astrophysical importance of the $^{22}$Ne($\alpha,n$)$^{25}$Mg reaction, it has been the focus of a considerable number of studies  \cite{GIESEN199395,PhysRevLett.87.202501,PhysRevC.80.055803,PhysRevC.82.025802,PhysRevC.76.025802,PhysRevC.85.044615,PhysRevC.93.055803,PhysRevC.66.055805,PhysRevC.85.065809}. Direct measurements of the $^{22}$Ne($\alpha,n$)$^{25}$Mg reaction have been carried out down to $E_r = 832$ keV ($E_x = 11.319$ MeV) \cite{PhysRevLett.87.202501}. For resonances lower than this, various indirect methods - briefly summarised below - have been used to try to constrain the $^{22}$Ne($\alpha,\gamma$)$^{26}$Mg and $^{22}$Ne($\alpha,n$)$^{25}$Mg reaction rates.

Longland \textit{et al.} \cite{PhysRevC.80.055803} used the inelastic scattering of polarised $\gamma$ rays - denoted as $^{26}$Mg($\overline{\gamma},\gamma^\prime$)$^{26}$Mg - to assign $J^\pi$s to levels in $^{26}$Mg. This technique is extremely powerful as it allows for clear and incontrovertible discrimination between $1^-$ and $1^+$ states. This reaction is, however, unable to populate $0^+$ states due to the $\gamma$-ray angular momentum selection rules.

Talwar \textit{et al.} \cite{PhysRevC.93.055803} used the $^{26}$Mg($\alpha,\alpha^\prime$)$^{26}$Mg reaction to populate states in $^{26}$Mg. This reaction preferentially populates low-spin, natural-parity states with the same isospin ($T=1$) as the ground state of $^{26}$Mg - the states which will contribute to the $^{22}$Ne$+\alpha$ reactions. The high level density can make it difficult to identify states clearly, however. The shapes of the differential cross sections from these reactions allow for assignment of spin and parity to be made.

Talwar \textit{et al.} \cite{PhysRevC.93.055803} and others \cite{PhysRevC.76.025802,GIESEN199395,ota20146li} have used the $\alpha$-particle transfer reaction $^{22}$Ne($^6\mathrm{Li},d$)$^{26}$Mg to attempt to determine spins, parities and $\alpha$-particle partial widths of states in $^{26}$Mg. While this reaction may be used to estimate the $\alpha$-particle partial width, the contribution of other reaction mechanisms such as incomplete fusion and multi-step reactions can make the extraction of the $\alpha$-particle partial width subject to considerable theoretical uncertainties unless the asymptotic normalisation coefficient can be extracted \cite{PhysRevC.90.042801}.

The $^{25}$Mg($n,\gamma$)$^{26}$Mg reaction has been studied using the n-TOF facility at CERN \cite{PhysRevC.85.044615} and the Oak Ridge Electron Linear Accelerator \cite{PhysRevC.66.055805}. In these experiments, several resonances above the neutron threshold were identified. A number of properties of these resonances (resonance strengths, spins and parities) were measured or constrained, improving the estimate of the $^{22}$Ne($\alpha,n$)$^{25}$Mg reaction. As the states populated in this reaction are above the neutron threshold, few constraints can be easily provided for the contribution to the $^{22}$Ne($\alpha,\gamma$)$^{26}$Mg reaction from states below the neutron threshold. 

In addition to these studies, which have focussed on the astrophysical states of interest, a measurement of the $^{26}$Mg($p,p^\prime$)$^{26}$Mg reaction using a 20-MeV proton beam and a Q3D magnetic spectrometer has been performed \cite{Moss1976429}. The number and energies of states in $^{26}$Mg were measured up to $E_x = 11.171$ MeV, but no information on spins and parities of the states was obtained.

In this paper, we report on a measurement of the $^{26}$Mg($\alpha,\alpha^\prime$)$^{26}$Mg reaction using the K600 magnetic spectrometer at iThemba LABS, South Africa. This measurement was taken as part of an on-going series of studies \cite{PhysRevC.95.024319} searching for monopole and dipole states as signatures of clustering in light nuclei (see e.g. \cite{Kawabata20076,PhysRevC.93.034319}). The experiment is similar to the measurement of Talwar \textit{et al.} \cite{PhysRevC.93.055803}, although some differences in the interpretation of the two experiments have been found. As such, we have attempted to evaluate the results of the present experiment, highlighting the discrepancies between this experiment and Ref. \cite{PhysRevC.93.055803}, as well as other experimental studies, to try to provide a consistent description of the states which contribute to the $^{22}$Ne($\alpha,n$)$^{25}$Mg and $^{22}$Ne($\alpha,\gamma$)$^{26}$Mg reactions.

\section{Experimental Method}

The experimental method is identical to that described in Ref. \cite{PhysRevC.95.024319}. A 200-MeV dispersion-matched $\alpha$-particle beam was scattered off an enriched $^{26}$Mg target. Scattered particles were momentum-analysed in the K600 Q2D magnetic spectrometer \cite{Neveling201129}. A plastic scintillator at the focal plane was used to trigger the data acquisition and to measure the energy deposited by the particle hitting the focal plane. The time between the focal-plane hit and the accelerator RF pulse was recorded, giving a measure of the time-of-flight of the scattered particle through the spectrometer. Particle identification was carried out using the energy deposited at the focal plane and the time-of-flight through the spectrometer. Particle positions and trajectories at the focal plane were measured using two vertical drift chambers.

The data were acquired in two experiments. The first experiment was in the 0-degree mode of the K600 \cite{Neveling201129} using a $99.94\%$ isotopically enriched $^{26}$Mg target of areal density $1.33$ mg/cm$^2$. In this mode, the unreacted beam passed through the spectrometer before being stopped on a Faraday cup located within the wall of the vault. An unavoidable background was observed resulting from particles which had scattered off the target and the inside of the spectrometer. In order to quantify and subtract this background, the spectrometer was operated in focus mode in which the reaction products were vertically focussed by the spectrometer quadrupole onto a vertically narrow band on the focal plane. Using a well-established method \cite{Neveling201129}, the off-focus regions of the focal plane was used to construct background spectra which were subtracted from the in-focus region to produce background-subtracted spectra. However, because the reaction products were focussed vertically, any information on the vertical scattering angle was lost. This limits the data to a cross section for the full acceptance of $\theta_{\mathrm{lab}}<2$\textdegree.

The second experiment was performed in the small-angle mode of the K600 in which the spectrometer aperture was placed at $\theta_{\mathrm{lab}}=4$\textdegree, covering $\theta_{\mathrm{lab}}=2$\textdegree$- 6$\textdegree. In this experiment, the target had $99.94\%$ enrichment, and an areal density of $0.6$ mg/cm$^2$. In the small-angle mode, the unreacted beam was stopped on a Faraday cup adjacent to the spectrometer aperture just before the spectrometer quadrupole. In this mode, because the target-induced background was much lower, the spectrometer was operated in under-focus mode, maintaining the link between the vertical focal-plane position and the vertical scattering angle. The scattering angle of the scattered particle was reconstructed from the vertical focal-plane position and the horizontal trajectory. The dependence of the scattering angles on the focal-plane position and trajectory was calibrated using a multi-hole collimator \cite{Neveling201129}.

The focal plane was calibrated using well-known states in $^{24}$Mg. A linear offset was introduced into the focal-plane excitation energy to account for the differing target thickness as the energy loss for targets scattered from the front and back of the target are almost identical. The calibrations are validated by confirming that well-known states in $^{26}$Mg (the $1^-$ states at 10.495 and 10.575 MeV which have been observed in $^{26}$Mg($\gamma,\gamma^\prime$)$^{26}$Mg reactions \cite{PhysRevC.80.055803}) appear at the known excitation energy. The difference between the weighted means of these level energies and the known energies of these levels was 5 keV and this was taken as the systematic uncertainty. This was combined with the statistical uncertainty in the peak-fitting to give a total uncertainty.

\begin{figure*}
\includegraphics[width=\textwidth]{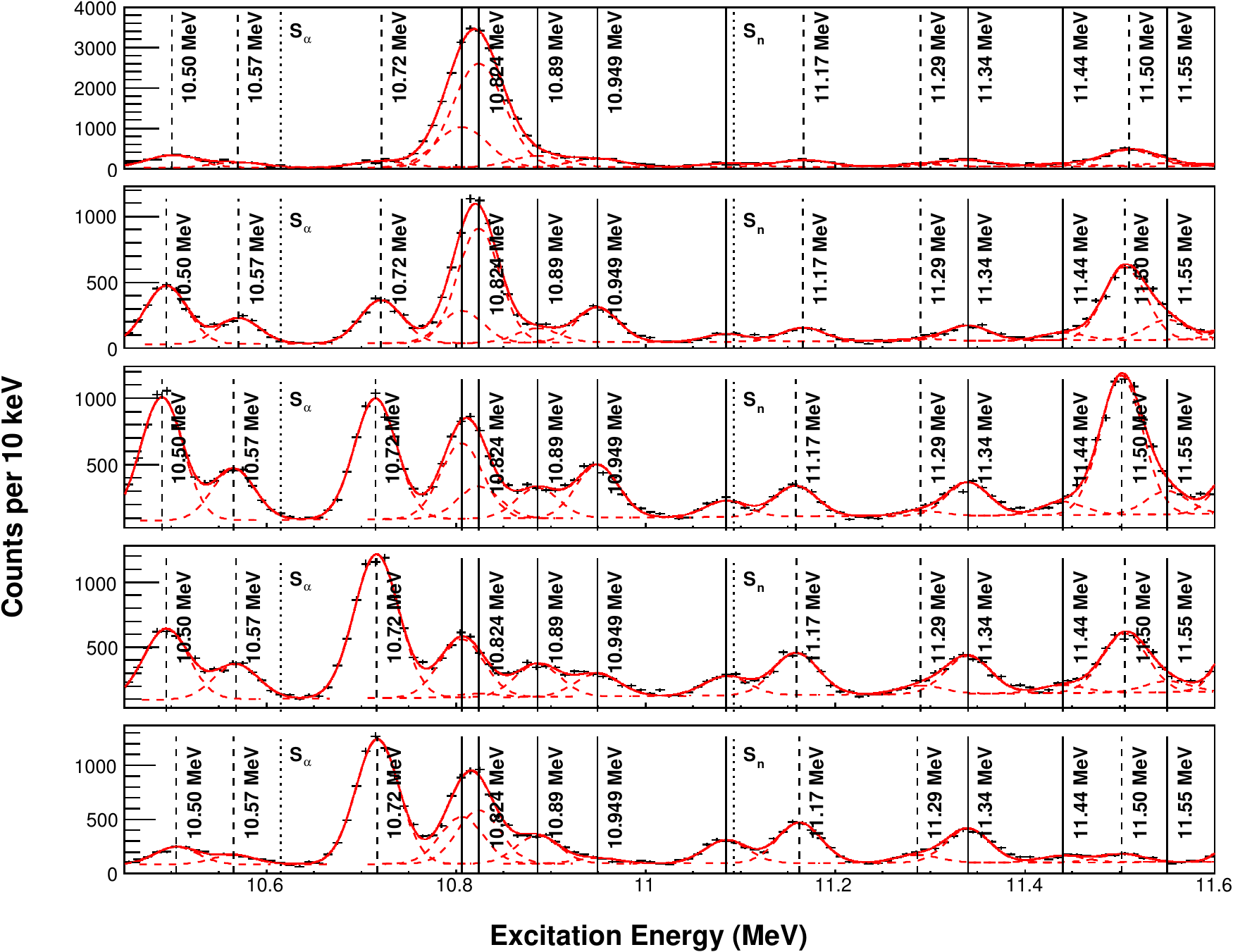}
 \caption{Excitation-energy spectra for the $^{26}$Mg($\alpha,\alpha^\prime$)$^{26}$Mg reaction for the $\theta_{\mathrm{lab}}<$2\textdegree\ angle bite (top), 2\textdegree$<\theta_{\mathrm{lab}}<$3\textdegree\ (second), 3\textdegree$<\theta_{\mathrm{lab}}<$4\textdegree (middle), 4\textdegree$<\theta_{\mathrm{lab}}<$5\textdegree\ (fourth) and 5\textdegree$<\theta_{\mathrm{lab}}<$6\textdegree (bottom). The solid vertical lines are states with fixed energies, the dashed vertical lines are states without fixed energies and the dotted lines show the positions of the $\alpha$-particle and neutron thresholds. The solid red line is the total fit and the dashed red lines are the contributions from individual peaks.}
 \label{fig:spectra1}
\end{figure*}

Excitation-energy spectra were fitted with a number of Gaussian peaks in order to identify states and extract differential cross sections (Figure \ref{fig:spectra1}). The Gaussians have a common width representing the experimental resolution of 64 keV FWHM for the 0-degree data and 53 keV FWHM for the small-angle data. The 0-degree spectrum had a background subtraction performed and was thus fitted using the $\chi^2$ method. The other spectra were fitted using the log-likelihood method. Owing to the high level density relative to the experimental resolution, it was necessary to fix states with known excitation energy e.g. the states observed in Ref. \cite{PhysRevC.80.055803}. Other states, which are cleanly observed at some angles and not others, were also fixed based on the fits for those angle bites, see Table \ref{tab:Mg26_Jpis} for details. The region of the fit was limited to just below the $\alpha$-particle threshold ($E_x = 10.615$ MeV) up to $E_x = 11.6$ MeV which covers the Gamow window for the $^{22}$Ne$+\alpha$ reactions ($E_x = 10.85 - 11.5$ MeV) at astrophysically relevant temperatures \cite{PhysRevC.85.065809}. The reason that the analysis included the region below the $\alpha$-particle threshold was to make use of the previously identified 10.577-MeV $1^-$ state observed in $^{26}$Mg($\overline{\gamma},\gamma^\prime$)$^{26}$Mg \cite{PhysRevC.80.055803}, which was used for comparison with other potential $1^-$ states.

\begin{figure*}
 \includegraphics[width=\textwidth]{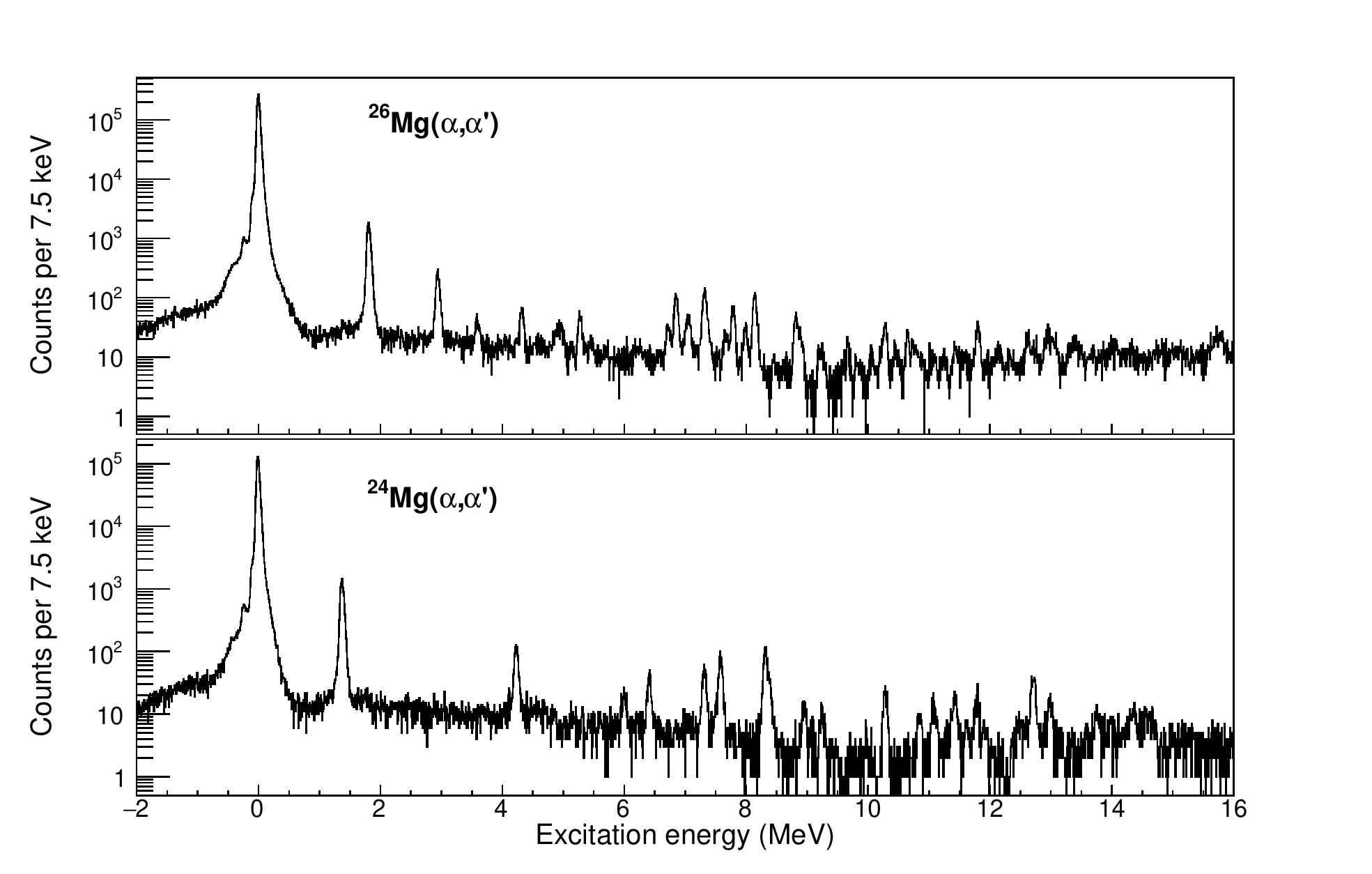}
 \caption{Excitation-energy spectra taken with the field setting that includes the elastic peak with the spectrometer aperture centred on $\theta = 6$\textdegree. The top spectrum is $^{26}$Mg data and the bottom spectrum is $^{24}$Mg data. The broad background up to around 9 MeV is due to $p(\alpha,\alpha^\prime)p$ reactions off water in the target.}
 \label{fig:elastics}
\end{figure*}

In order to verify that the $^{26}$Mg data were free from contamination from other magnesium isotopes, data were also taken covering a lower excitation-energy region, including the ground state, during the small angle experiment. The first $2^+$ state from $^{24}$Mg (at $E_x = 1369$ keV) is not observed in these data which leads us to conclude that none of the states observed in the present experiment are likely to result from $^{24}$Mg. The equivalent spectra for $^{24}$Mg and $^{26}$Mg are shown in Figure \ref{fig:elastics}. The other major target contaminants ($^{12}$C and $^{16}$O) are not observed strongly in these data. In addition, previous experimental studies of the $^{12}$C($\alpha,\alpha^\prime$)$^{12}$C and $^{16}$O($\alpha,\alpha^\prime$)$^{16}$O reactions have shown that there is only one narrow state in the excitation-energy region discussed in this paper, a $J^\pi = 2^+$ state in $^{16}$O at 11.52 MeV \cite{PhysRevC.95.031302}. Small amounts of water are present in the target and are responsible, through the $p(\alpha,\alpha)p$ reaction, for the broad structure at lower excitation energies. Based on these considerations we conclude that the states observed in the present experiment all originate from $^{26}$Mg.

\begin{figure*}
\includegraphics[width=\textwidth]{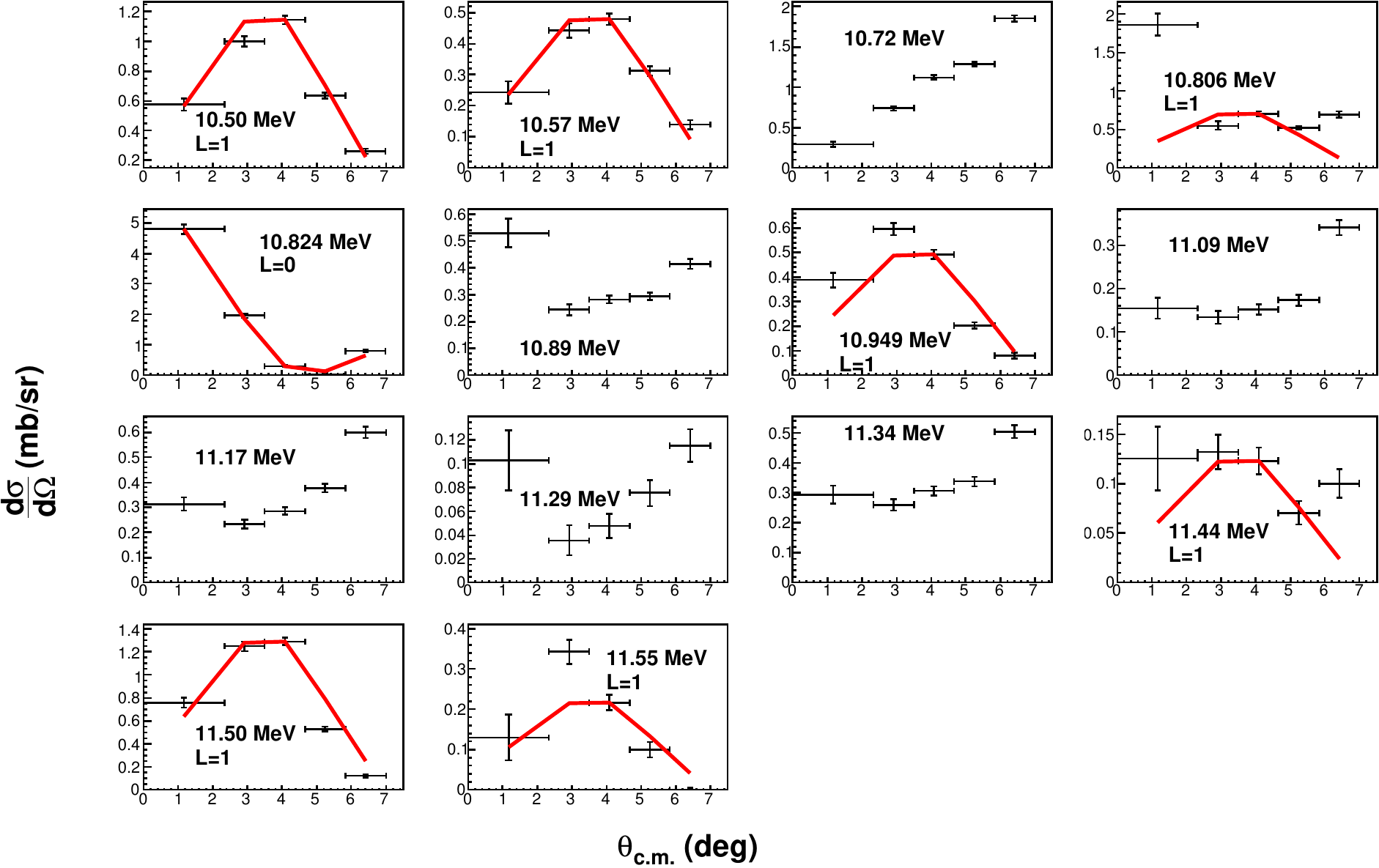}
 \caption{Experimental differential cross sections for various states in $^{26}$Mg. The states are labelled on the corresponding figure. The red solid lines are the angle-averaged DWBA differential cross sections for that particular state, normalised to the 0-degree datum for $\ell=0$ transitions, and to the $\theta_{c.m.} = 3-4$ degree datum for $\ell=1$ transitions. For more detailed discussions of the differential cross sections, consult the text.}
 \label{fig:differentialCSs}
\end{figure*}

The differential cross sections (Figure \ref{fig:differentialCSs}) were used to make assignments of the $\ell$-value of the reaction. As the focus of the original experiment was on monopole and dipole states, only the 0\textdegree-6\textdegree\ laboratory scattering angle region was covered, and thus only $\ell=0$ or $\ell=1$ assignments can be made. It is helpful to set out qualitatively the shapes of the differential cross sections in this experiment. The signature of the differential cross sections can then be followed in the experimental excitation energy spectra (Figure \ref{fig:spectra1}) at different angles in addition to the differential cross sections generated by fitting the spectra. For $\ell=0$ transitions, the differential cross section shows a strong peak at a scattering angle of $\theta_{\mathrm{lab}}=0$\textdegree\ and a minimum around $\theta_{\mathrm{lab}}=4$\textdegree. For $\ell=1$ transitions, the signature of the differential cross section is less obvious - the differential cross section has a peak between $\theta_{\mathrm{lab}}=3$\textdegree and $4$\textdegree\ and drops off strongly towards $\theta_{\mathrm{lab}}=6$\textdegree, the edge of the aperture for the small-angle measurement. These behaviours are well understood and have been observed previously in a similar experiment using $^{24}$Mg and $^{28}$Si \cite{PhysRevC.95.024319}.

The differential cross sections were compared to Distorted-Wave Born Approximation (DWBA) calculations. The optical-model potential used for the DWBA calculations is that of Nolte, Machner and Bojowald \cite{PhysRevC.36.1312}. The optical-model parameters for the Woods-Saxon potential are given in Table \ref{tab:OMPParams}. To ensure that the DWBA calculations can be directly compared to the data (i.e. that the calculated quantity resulting from the DWBA calculations is the same as the experimental observable), the DWBA differential cross sections were integrated over the pertinent angular regimes, and the effective differential cross section for that angular regime was computed. The DWBA differential cross sections were only computed for $\ell=0$ and $\ell=1$ transitions as these were the only angular momentum values which may be firmly assigned in the present experiment.

\begin{table}
\caption{The optical model potential parameters used for the DWBA calculations. $V$ and $W$ are the real and imaginary potential depths respectively. $r_{0R}$ ($r_{0I}$) is the reduced radius for the real (imaginary) part of the potential and $a_{R}$ ($a_{I}$) is the diffusivity. The reduced Coulomb radius is given by $r_{0c}$.}
\begin{tabular}{c|c}
  Parameter & Value \\
  \hline \hline
  $V$ & 76.01 MeV \\
  $r_{0R}$ & 1.245 fm \\
  $a_{R}$ & 0.79 fm \\
  $W$ & 22.97 MeV \\
  $r_{0I}$ & 1.57 fm \\
  $a_{I}$ & 0.63 fm \\
  $r_{0c}$ & 1.3 fm
\end{tabular}
\label{tab:OMPParams}
\end{table}

\begin{table*}
\caption{Energy levels and $J^\pi$ assignments from various sources for $^{26}$Mg levels. The adopted values for the excitation energy and the $J^\pi$ are given in the final two columns.}
 \begin{tabular}{c|c||c|c||c|c||c|c||c|c||c|c}
  $E_x$ [MeV] \footnote{Present experiment} & $J^\pi$ $^\mathrm{a}$ & $E_x$ [MeV] \footnote{Ref. \cite{PhysRevC.93.055803}: $^{26}$Mg($\alpha,\alpha^\prime$)$^{26}$Mg} & $J^\pi$ $^\mathrm{b}$ & $E_x$ [MeV] \footnote{Ref. \cite{PhysRevC.80.055803}: $^{26}$Mg($\gamma,\gamma^\prime$)$^{26}$Mg.} & $J^\pi$ $^\mathrm{c}$ & $E_x$ [MeV] \footnote{Ref. \cite{PhysRevC.85.044615}: $^{25}$Mg($n,\gamma$)$^{26}$Mg.} & $J^\pi$ $^\mathrm{d}$ & $E_x$ [MeV] \footnote{Ref. \cite{PhysRevLett.87.202501}: $^{22}$Ne($\alpha,n$)$^{25}$Mg.} & $J^\pi$ $^\mathrm{e}$ & $E_x$ [MeV] \footnote{Adopted in the present work.} & $J^\pi$ $^\mathrm{f}$\\
  \hline \hline
  10.50(2) & $1^-$ & 10.495(9) & $1^-$ & & & & & & & 10.495(9) & $1^-$  \\
  10.57(1) & $1^-$ & 10.575(10) & $1^-, 2^+$ & 10.5733(8) & $1^-$ & & & & & 10.5733(8) & $1^-$\\
  10.72(1) & & 10.717(9) & $1^-, 2^+$ & & & & & & & 10.717(9) & $2^+$ \\
  10.806(10)\footnote{Fixed peak position - uncertainty is assumed to be 10 keV.} & & & & 10.8057(7) & $1^-$ & & & & & 10.8057(7) & $1^-$ \\
  10.824(10)$^\mathrm{g j}$ & $0^+$ & 10.822(10) & $1^-$ & & & & & & & 10.824(3)\footnote{$E_x$ value taken from Ref. \cite{Moss1976429}.} & $0^+$\\
  10.89(1)$^\mathrm{g}$\footnote{New state}  & $>1$ & & & & & & & & & 10.89(1) & $>1$ \\
  10.949(10)$^\mathrm{g}$ & $1^-$ & 10.951(21) & $1^-, 2^+$ & 10.9491(8) & $1^-$ & & & & & 10.9491(8) & $1^-$ \\
  11.085(10)$^\mathrm{g}$ & & 11.085(8) & $2^+, 3^-$ & & & & & & & 11.085(8) & $2^+, 3^-$ \\
  11.17(1) & & 11.167(8) & $1^-, 2^+$ & & & & & & & 11.167(8) & $(2^+)$ \\
  11.29(3) & $>1$ & 11.301(9) & & & & & & & & 11.301(9) & $>1$ \\
  11.34(2) & $>1$ & & & & & 11.3347(4) & $(1^-)$ & & & 11.3347(4) & $>1$ \\
  11.44(1)$^\mathrm{g}$ & & 11.445(9) & $1^-, 3^-$ & & & & & 11.441(2) & & 11.441(2) & $1^-, 3^-$ \\
  11.50(1)  & $1^-$ & 11.506(11) & $0^+, 1^-$ & & & 11.5001(4) & $(1^-)$ & 11.506(2) &  & 11.5001(4) & $1^-$ \\
  11.55(3)$^\mathrm{g j}$\footnote{State with changed assignment} & $(1^-)$ & & & & & & & & & (11.526) & $(1^-)$ \\
 \end{tabular}
\label{tab:Mg26_Jpis}
\end{table*}

\section{Discussion}

Based on the differential cross sections measured in the present experiment combined with other experimental data \cite{PhysRevC.93.055803,PhysRevC.80.055803}, we assign spins and parities to states as summarised in Table \ref{tab:Mg26_Jpis}. The rationales and further details for the assignments of certain states are set out below.

{\it 10.50 MeV:} This state is $J^\pi = 1^-$. It shows clear similarities with the known $1^-$ state at 10.573 MeV observed in $^{26}$Mg($\overline{\gamma},\gamma^\prime$)$^{26}$Mg measurements \cite{PhysRevC.80.055803}. This assignment is in agreement with that of Ref. \cite{PhysRevC.93.055803}.

{\it 10.57 MeV:} This state is the known $1^-$ state observed in $^{26}$Mg($\overline{\gamma},\gamma^\prime$)$^{26}$Mg measurements \cite{PhysRevC.80.055803}. It shows a clear $\ell=1$ differential cross section in good agreement with the DWBA calculations and previously observed $J^\pi = 1^-$ states in $^{28}$Si \cite{PhysRevC.95.024319}.

{\it 10.72 MeV:} We favour an assignment of $J>1$. The possible values for this state are $1^-$ and $2^+$ \cite{PhysRevC.93.055803}. Excluding $J^\pi = 1^-$ results in an assignment of  $J^\pi = 2^+$. This state does not appear in the calculation of the reaction rates by Longland, Iliadis and Karakas \cite{PhysRevC.85.065809}.

{\it 10.806 MeV:} This is the $J^\pi=1^-$ state observed by Longland {\it et al.} \cite{PhysRevC.80.055803}, and is not well resolved from the reassigned 10.824-MeV $J^\pi = 0^+$ state; this is likely the reason for the high cross section observed at $\theta_{\mathrm{lab}}<2^\circ$. This results in a differential cross section which does not show a clear $\ell=1$ pattern. The $J^\pi = 1^-$ assignment given by Longland {\it et al.} is, however, conclusive.

{\it 10.824 MeV:} This state shows a clear $\ell=0$ distribution and must have $J^\pi = 0^+$, and may be associated with the one observed by Goss at $E_x = 10.824(3)$ MeV \cite{Moss1976429}. While the rationale given in Ref. \cite{PhysRevC.93.055803} as to why the state observed in that experiment at $E_x = 10.822$ MeV cannot be the $J^\pi = 1^+$ state observed at $E_x = 10.81$ MeV in $^{26}$Mg($p,p^\prime$)$^{26}$Mg experiments - that unnatural-parity states are not strongly populated in inelastic $\alpha$-particle scattering - is correct, based on the differential cross section this state cannot be the $J^\pi = 1^-$ state observed in $^{26}$Mg($\overline{\gamma},\gamma^\prime$)$^{26}$Mg measurements \cite{PhysRevC.80.055803}. This leads to the conclusion that there is a third state at this energy with $J^\pi = 0^+$. In conclusion: there are three states at approximately this energy: a $J^\pi = 1^+$ state at 10.81 MeV \cite{PhysRevC.39.311}, the 10.806-MeV $J^\pi = 1^-$ state \cite{Moss1976429} and a 10.824-MeV $J^\pi = 0^+$ state observed in $^{26}$Mg($\alpha,\alpha^\prime$)$^{26}$Mg (present experiment). We note that the differential cross section shown in Ref. \cite{PhysRevC.93.055803} is not inconsistent with a $J^\pi=0^+$ assignment.

With the reassignment of a different state at this energy, it is not clear if the previously accepted association (e.g. \cite{PhysRevC.85.065809}) between the $J^\pi = 1^-$ 10.806-MeV state observed in $^{26}$Mg($\overline{\gamma},\gamma^\prime$)$^{26}$Mg \cite{PhysRevC.80.055803} and the 10.808(20)-MeV state observed in $^{22}$Ne($^6\mathrm{Li},d$)$^{26}$Mg \cite{PhysRevC.76.025802} still holds.


In addition, we note that an additional $J^\pi = 2^+$ state has been identified at $E_x = 10.838(24)$ MeV in $^{26}$Mg($e,e^\prime$)$^{26}$Mg reactions, and that Moss made a connection between this state and the 10.824-MeV state observed in that experiment. Based on the present reassignments, we do not observe a $J^\pi = 2^+$ state as observed in Ref. \cite{0301-0015-7-8-005} meaning that another unobserved state may exist at around this excitation energy.

{\it 10.89 MeV:} This state is not reported in Ref. \cite{PhysRevC.93.055803} and there is no obvious correspondence between this state and any of those listed in Ref. \cite{PhysRevC.85.065809}. However, the state is clearly observed in the present experiment, especially in the $4$\textdegree$<\theta_{\mathrm{lab}}<5$\textdegree\ and $5$\textdegree$<\theta_{\mathrm{lab}}<6$\textdegree\ angle bites. The higher cross section observed at $\theta_{\mathrm{lab}}<2$\textdegree\ is due to the strength of the unresolved $J^\pi = 0^+$ state at 10.824 MeV. This state is tentatively given an assignment of $J>1$ because the largest cross section is observed in $4$\textdegree$<\theta_{\mathrm{lab}}<5$\textdegree\ and $5$\textdegree$<\theta_{\mathrm{lab}}<6$\textdegree\ angle bites. This state may correspond to the 10.881- or 10.893-MeV states observed by Moss \cite{Moss1976429}.

{\it 10.949 MeV:} Based on the differential cross section observed, we concur with the $J^\pi=1^-$ assignment of Ref. \cite{PhysRevC.93.055803}. This state is the 10.949 MeV $J^\pi = 1^-$ state listed in Ref. \cite{PhysRevC.85.065809} which was observed in $^{26}$Mg($\overline{\gamma},\gamma^\prime$)$^{26}$Mg \cite{PhysRevC.80.055803} and $^{26}$Mg($p,p^\prime$)$^{26}$Mg \cite{Moss1976429}.

{\it 11.085 MeV:} This state is observed in Ref. \cite{PhysRevC.93.055803} and in the present experiment. The differential cross section is indicative of a spin of $J>1$. Owing to the limited angular range studied in the present experiment, we are unable to improve upon the assignment of $J^\pi = 2^+$ or $3^-$ given in Ref. \cite{PhysRevC.93.055803}.

{\it 11.17 MeV:} This state is observed in both Ref. \cite{PhysRevC.93.055803} and the present experiment. The differential cross section does not support an assignment of $J^\pi = 1^-$. Ref. \cite{PhysRevC.93.055803} limited the potential $J^\pi$ to $1^-$ or $2^+$. A tentative assignment of $J^\pi=2^+$ is reasonable, in the absence of a definitive differential cross section. As such, it is reasonable to associate this state with the 11.163-MeV, $J^\pi = 2^+$ state seen in $^{25}$Mg($n,\gamma$)$^{26}$Mg reactions \cite{PhysRevC.85.044615}.

{\it 11.29 MeV:} This state is observed in Ref. \cite{PhysRevC.93.055803}. The differential cross section does not show any clear $\ell=0$ or $\ell=1$ shape leading to a conclusion that this state has $J>1$. A number of states at this approximate excitation energy have been observed in $^{25}$Mg($n,\gamma$)$^{26}$Mg reactions. It is unclear which of these states corresponds to the state observed in the present experiment.

{\it 11.34 MeV:} This state may correspond to the 11.335-MeV state observed in Refs. \cite{PhysRevC.66.055805,PhysRevC.85.044615} at $E_n = 253$ keV. This resonance is tentatively assigned $J^\pi = 1^-$ in those studies. On the basis of the present experiment, in which the differential cross section does not have an $\ell=1$ shape, an assignment of $J>1$ is given.

{\it 11.44 MeV:} This state is not well resolved from the $1^-$ state at 11.50 MeV. The potential assignments made in Ref. \cite{PhysRevC.93.055803} are $J^\pi = 1^-$ or $3^-$. The differential cross section is not conclusive. This state may be the 11.441-MeV state observed in $^{22}$Ne($\alpha,n$)$^{25}$Mg reactions \cite{PhysRevLett.87.202501}.

{\it 11.50 MeV:} The state observed at 11.50 MeV was also observed in Ref. \cite{PhysRevC.93.055803}. This state is probably the 11.506-MeV resonance observed in $^{22}$Ne($\alpha,n$)$^{25}$Mg \cite{PhysRevLett.87.202501} and the 11.50-MeV ($E_n = 423.43$ keV) resonance observed in $^{25}$Mg($n,\gamma$)$^{26}$Mg \cite{PhysRevC.85.044615}. From the differential cross section, this state must have $J^\pi = 1^-$. There is a known state in $^{16}$O at 11.52 MeV with $J^\pi = 2^+$ \cite{PhysRevC.95.031302} but both from the differential cross section in the present data, and from the discussion of possible target contaminants above, it is clear that the state observed in this experiment is distinct from the $^{16}$O state.

{\it 11.55 MeV:} This state is not well resolved from the 11.50-MeV $J^\pi =  1^-$ state. There is no obvious corresponding known state for this state. There is, however, a resonance reported in the literature at 11.526 MeV which has $J^\pi = 1^-$ \cite{PhysRevC.85.065809}. It is possible that these states are, in fact, the same state and that the shift is an artefact of the inability to resolve the 11.50- and 11.55-MeV states. For the purposes of calculating the rate below we use the data derived from direct measurements where available \cite{PhysRevLett.87.202501} and so this state is omitted.

\section{Implications for the astrophysical $^{22}$Ne($\alpha,\gamma$)$^{26}$Mg reaction rate}

The $^{22}$Ne($\alpha,n$)$^{25}$Mg reaction can contribute as a neutron source both in the advanced thermal pulses in AGB stars and also in the weak $s$-process in core-helium burning in massive stars. In the case of core-helium burning, the contribution of the $^{22}$Ne($\alpha,n$)$^{25}$Mg is negligible until the end of helium burning when the temperature exceeds $0.25$ GK. Before this point, $^{22}$Ne may be depleted by the $^{22}$Ne($\alpha,\gamma$)$^{26}$Mg reaction which reduces the final neutron exposure achieved during the final stages of core-helium burning. Thus, the behaviour of the $^{22}$Ne($\alpha,\gamma$)$^{26}$Mg reaction at lower temperatures ($T_9 < 0.3$ GK) can influence the final $s$-process production.

In this paper, the particular changes that have been made to the assignments of levels in $^{26}$Mg are: a $0^+$ state at 10.82 MeV has been observed in addition to the $1^-$ state at 10.805 MeV, and a new state with $J>1$ has been observed at 10.89 MeV. In addition to the reassignment and the new 10.89-MeV state, the 10.838-MeV $J^\pi=2^+$ state observed in $^{26}$Mg($e,e^\prime$)$^{26}$Mg has been included as it has previously been omitted from calculations of the rate, but cannot be firmly associated with another state in $^{26}$Mg such as the 10.824-MeV state which has a different $J^\pi$. The 11.29-MeV state has been omitted as it probably corresponds to one of a number of known states at around this excitation energy \cite{PhysRevC.85.065809} which are included within the STARLIB calculation already. Due to its non-observation in direct measurements, the potential new $E_x = 11.55$-MeV state is not included in the STARLIB calculation as it has not been observed in direct measurements of the $^{22}$Ne($\alpha,n$)$^{25}$Mg reaction \cite{PhysRevLett.87.202501}.

To quantify the effect of the changes to the resonances made in this experiment, a Monte Carlo simulation using the STARLIB tool \cite{ILIADIS2010251} has been performed. Input information for most resonances has been taken as the STARLIB default with only those resonances listed above changing. The 10.89-MeV state calculation has been performed twice, assuming $J^\pi = 2^+$ and $J^\pi = 3^-$ respectively in the two cases. The input information is summarised in Table \ref{tab:MCSimulationInput} and the input files themselves are available in an online repository \cite{STARLIBRepo}.

\begin{table}
 \caption{State information for the resonances in $^{26}$Mg used for the Monte-Carlo calculation of the $^{22}$Ne($\alpha,\gamma$)$^{26}$Mg reaction rate.}
 \begin{tabular}{| c | c | c | c |}
  $E_x$ [MeV] & $E_{r, c.m.}$ [keV] & $J^\pi$ & $\Gamma_{\alpha,sp}$ [eV] \\
  \hline \hline
  10.717(9)  & 102.2(9) & $2^+$ & $1.01 \times 10^{-47}$ \\
  10.8057(7) & 191.9(7) & $1^-$ & $6.65 \times 10^{-22}$ \\
  10.824(3)  & 209(3)     & $0^+$ & $5.34 \times 10^{-20}$ \\
  10.838(24) & 223(24)    & $2^+$ & $7.76 \times 10^{-20}$ \\
  10.89(1)   & 275(10)  & $2^+$ & $1.62 \times 10^{-16}$ \\
             & 275(10)  & $3^-$ & $2.05 \times 10^{-17}$ \\
 \end{tabular}
 \label{tab:MCSimulationInput}
\end{table}

The $\alpha$-particle widths for the new resonances have been calculated using $\Gamma_\alpha = 2 P_\ell(E) \gamma_\alpha^2$ where the reduced width $\gamma_\alpha^2$ is given by \cite{PhysRevC.85.065809}:

\begin{align}
 \gamma_\alpha^2 &= \frac{\hbar^2}{\mu a^2}\theta_\alpha^2 \\
		 &= \frac{\hbar^2}{2\mu a}S_\alpha \phi^2(a).
\end{align}
The $P_\ell(E)$ factor is the R-matrix penetrability which is calculated using \cite{thompson2009nuclear}:

\begin{equation}
 P_\ell(E) = \frac{ka}{F_\ell(E)^2 + G_\ell(E)^2},
\end{equation}
where $F_\ell(E)$ and $G_\ell(E)$ are respectively the regular and irregular Coulomb functions \cite{abramowitz2012handbook}, $k$ is the wavenumber and $a = 1.25 (22^{1/3} + 4^{1/3})$ fm is the channel radius. 

For the purposes of quantifying the upper-limits of the resonance strengths, STARLIB samples the width according to a Porter-Thomas distribution \cite{ILIADIS2010251}. We have, in this case, assumed that the upper-limit for the width is that from the R-matrix calculation and that the dimensionless reduced widths are distributed according to a Porter-Thomas distribution with $\theta^2 = 0.01$. However, other authors \cite{PhysRevC.85.065809} have assumed a smaller upper limit on the $\alpha$-particle width for the 10.806-MeV state than has been assumed in the present calculation. The upper limit on this state has been relaxed as the correspondence between the states observed in $^{22}$Ne($^6$Li,$d$)$^{26}$Mg and those states observed using other reactions is now unclear.

\begin{figure}
 \includegraphics[height=0.5\textwidth,angle=270]{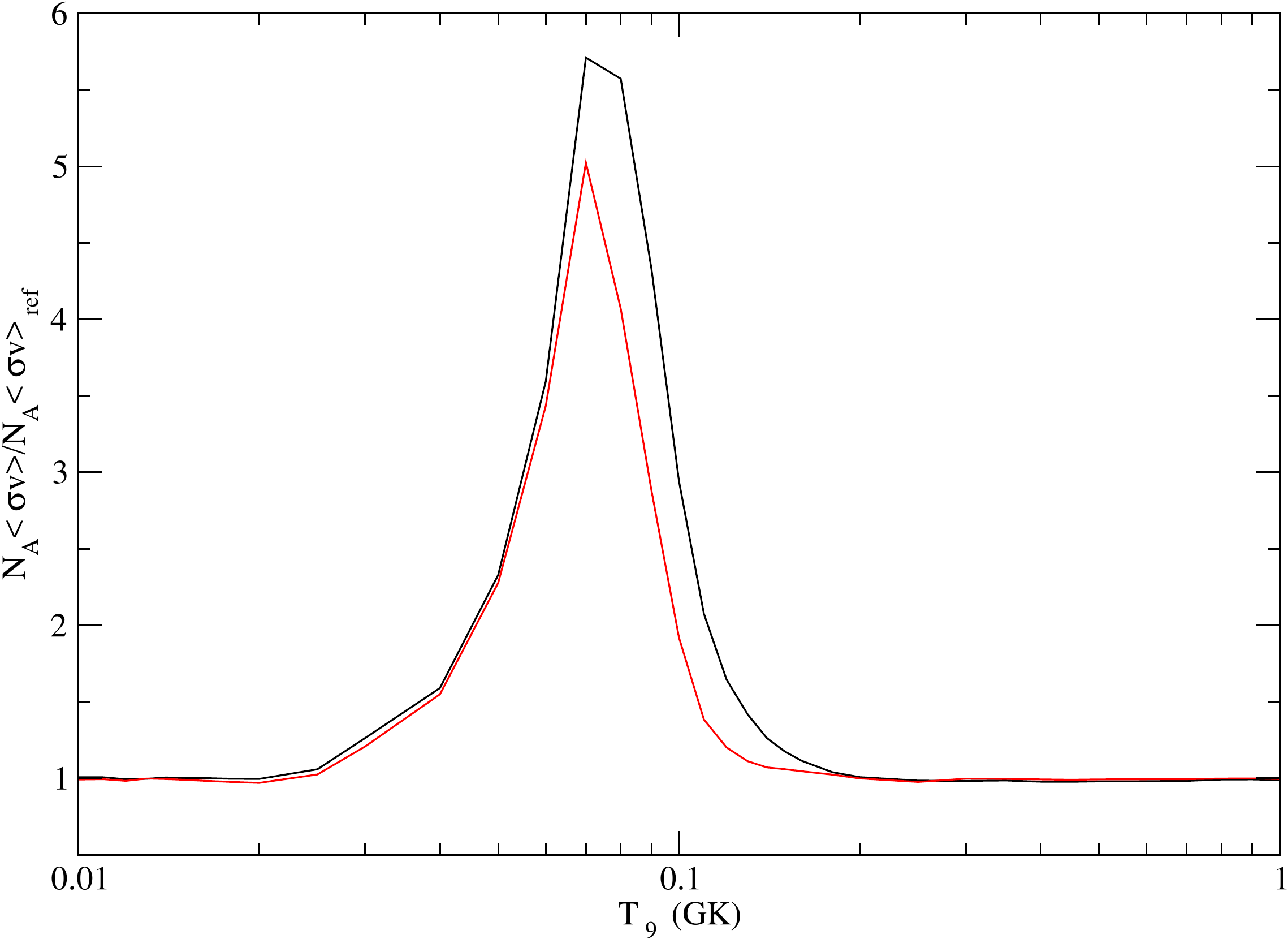}
 \caption{Ratio of the new to old median rates. For details of the calculation inputs, see the text. The black curve is the ratio with a $J^\pi=2^+$ assumption for the 10.89-MeV state while the red curve is the calculation assuming a $J^\pi=3^-$ assignment.}
 \label{fig:ratio_of_median_rates}
\end{figure}

\begin{figure}
\includegraphics[height=0.5\textwidth,angle=270]{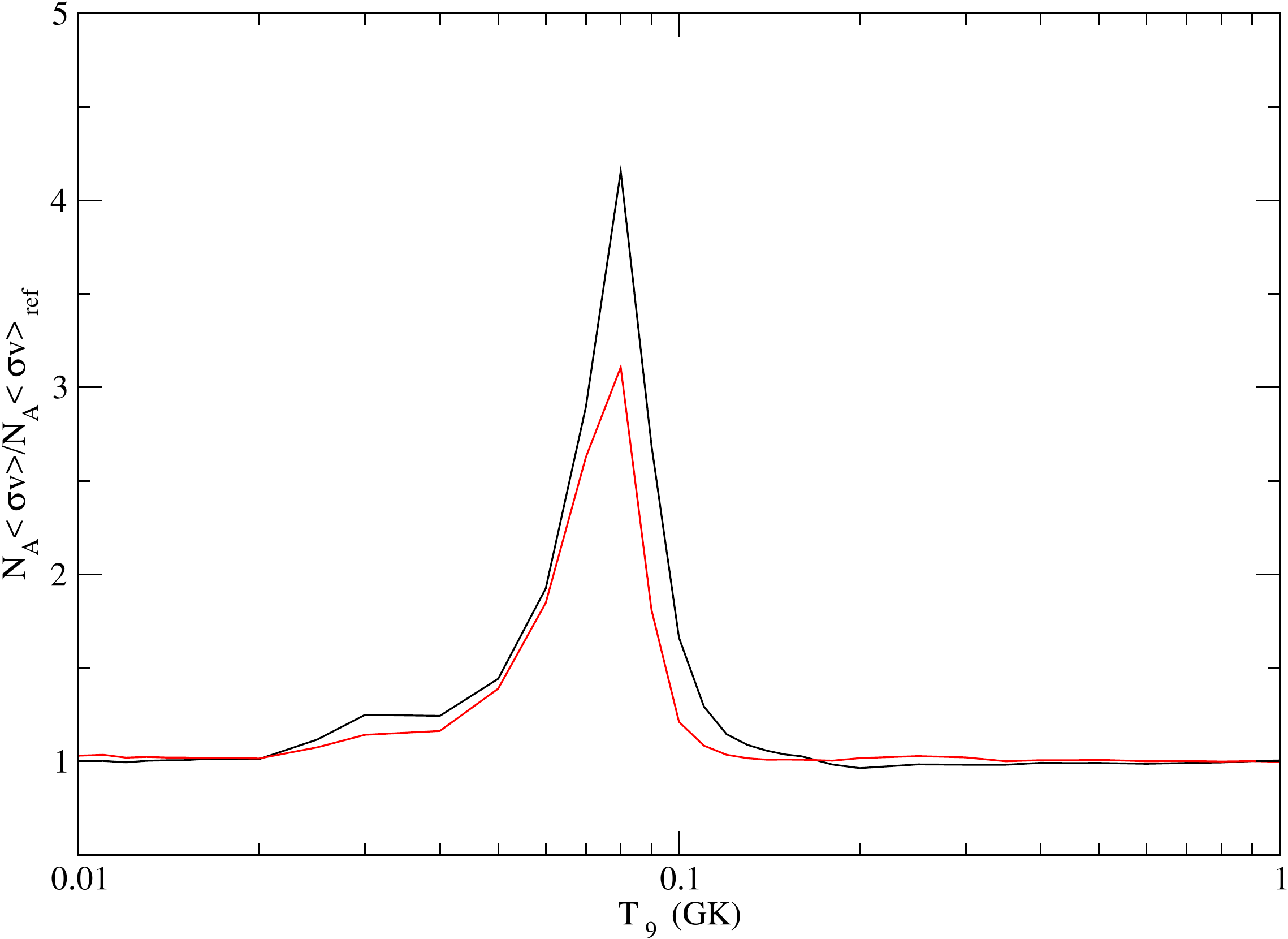}
 \caption{As Figure \ref{fig:ratio_of_median_rates} except for the ratios of the upper limits of the rates.}
 \label{fig:ratio_of_upper_limits}
\end{figure}

The ratios of the median rate of the current calculations to the STARLIB reference are shown in Figure \ref{fig:ratio_of_median_rates} and the ratios of the upper limits of the current calculations to the STARLIB reference are shown in Figure \ref{fig:ratio_of_upper_limits}. The new and reassigned resonances mainly contribute to an increase in the reaction rate at temperatures below $T_9 = 0.1$, as would be expected due to the low resonance energies of the reassigned resonances. This may increase consumption of $^{22}$Ne before the $^{22}$Ne($\alpha,n$)$^{25}$Mg reaction can operate.

\section{Conclusions}

The $^{26}$Mg($\alpha,\alpha^\prime$)$^{26}$Mg reaction was performed using the K600 magnetic spectrometer at iThemba LABS in South Africa. Spins and parities are deduced from the differential cross sections of scattered $\alpha$ particles and show some disagreement with the values deduced in a similar experimental study at RCNP Osaka \cite{PhysRevC.93.055803}. Reassigned states such as the now-separated $1^-$ state at 10.805 MeV and the $0^+$ state at 10.824 MeV, and a state with an unknown spin-parity at 10.89 MeV which is omitted from Refs. \cite{PhysRevC.85.065809} and \cite{PhysRevC.93.055803}. In addition, with the reassignment of the 10.822-MeV state from Ref. \cite{PhysRevC.93.055803}, there is evidence from a study of the $^{26}$Mg($e,e^\prime$)$^{26}$Mg reaction that another state at 10.838-MeV with $J^\pi = 2^+$ is present but not included in previous calculations of the reaction rate \cite{Moss1976429,0301-0015-7-8-005}. Additionally, $J^\pi = 1^-$ strength has been observed at $E_x = 11.55$ MeV; it is unclear if this is a new state or an artefact of the $E_x = 11.5$ and $E_x = 11.526$ MeV states known in this region \cite{PhysRevLett.87.202501}.

A new reaction rate and upper limit have been computed using the STARLIB tool to quantify the difference to the reaction rate caused by these additional states which shows that a considerable increase in the total reaction rate is possible at $T_9 = 0.04 - 0.1$.

It is clear from this paper that, despite the high number of experimental investigations of astrophysically important states in $^{26}$Mg, considerable uncertainty remains over the properties of the resonances in $^{26}$Mg including most fundamentally the number and excitation energies of the resonances. Part of this uncertainty arises from the highly selective reaction mechanisms used in many experimental studies of $^{26}$Mg which do not populate all of the states  coupled with energy resolutions which cannot resolve the energy levels. This is potentially a grave problem if, for example, measurements of $\alpha$-particle spectroscopic factors through $^{22}$Ne($^6$Li,$d$)$^{26}$Mg reactions cannot be firmly connected to particular known states due to insufficient excitation-energy resolution. More information on the number and position of levels in $^{26}$Mg is required, especially below the neutron threshold where $^{25}$Mg($n,\gamma$)$^{26}$Mg reactions cannot easily probe. Of particular use in separating the states below the neutron threshold in $^{26}$Mg may be coincidence measurements such as $^{26}$Mg($\alpha,\alpha\gamma$)$^{26}$Mg or $^{22}$Ne($^6$Li,$d\gamma$)$^{26}$Mg. 

\section{Acknowledgements}

The authors thank the accelerator group at iThemba LABS for the provision of a high-quality dispersion-matched $\alpha$-particle beam. PA thanks Drs R. Talwar and B. P. Kay for private communications relating to the RCNP data, and Mr Stuart Szwec for useful comments regarding optical model potentials. This work is based on the research supported in part by the
National Research Foundation of South Africa. RN acknowledges financial support from the NRF through grant number 85509.

\bibliography{Mg26_states}

\end{document}